\documentclass[letterpaper]{article}

\newif\iflong
\longtrue
\iflong
\usepackage[preprint]{spconf}
\else
\usepackage{spconf}
\fi

\usepackage{graphicx}
\usepackage{amsmath,amssymb,bm}
\usepackage{url}
\usepackage{setspace}
\usepackage[noadjust]{cite}

\usepackage{hyperref}
\usepackage[dvipsnames]{xcolor}
\newcommand\myshade{70}
\colorlet{mywholecolor}{MidnightBlue}
\hypersetup{
  linkcolor  = mywholecolor!\myshade!black,
  citecolor  = mywholecolor!\myshade!black,
  urlcolor   = mywholecolor!\myshade!black,
  colorlinks = true,
}

\newcommand{\ve}[1]{\textbf{#1}}

\usepackage{color}

\newcommand\minisection[1]{\vspace{1mm}\noindent\textbf{#1 ---}}

\title{SESQA: Semi-supervised Learning for Speech Quality Assessment}
\name{Joan Serr\`a, Jordi Pons, Santiago Pascual}
\address{
  Dolby Laboratories
  }
\begin{document}
\iflong\else
\ninept
\fi
\maketitle
\begin{abstract}
Automatic speech quality assessment is an important, transversal task whose progress is hampered by the scarcity of human annotations, poor generalization to unseen recording conditions, and a lack of flexibility of existing approaches. In this work, we tackle these problems with a semi-supervised learning approach, combining available annotations with programmatically generated data, and using 3~different optimization criteria together with 5~complementary auxiliary tasks. Our results show that such a semi-supervised approach can cut the error of existing methods by more than 36\%, while providing additional benefits in terms of reusable features or auxiliary outputs. Improvement is further corroborated with an out-of-sample test showing promising generalization capabilities.
\end{abstract}


%%%%%%%%%%%%%%%%%%%%%%%%%%%%%%%%%%%%%%%%%%%%%%%%%%%%%%%%%%%%%%%%%%%%%%%%%%%%%%%%%%%%%

\section{Introduction}
\label{sec:intro}

Speech quality assessment is crucial for a myriad of research topics and real-world applications. Its need ranges from algorithm evaluation and development to basic analytics or informed decision making. Speech quality assessment can be performed by subjective listening tests or by objective quality metrics~\cite{Loizou11BOOKCHAP}. Objective metrics that correlate well with human judgment open the possibility to scale up automatic quality assessment, with consistent results at a negligible fraction of the effort, time, and cost of their subjective counterparts. 

There have been considerable efforts in designing objective speech quality metrics. Traditional ones rely on standard signal processing blocks, like the short-time Fourier transform, or perceptually-motivated blocks, like the Gammatone filter bank. Together with further processing blocks, they conform an often intricate and complex rule-based system
~\cite{Rix01ICASSP, Kim05TASLP, Falk10TASLP, Malfait11TASL, Berends13JAES, chinen_visqol_2020}. Another, perhaps more recent alternative is to learn speech quality directly from raw data, by combining machine learning techniques with carefully chosen stimuli and their corresponding human ratings~\cite{Grancharov06TASLP, fu_qualitynet_2018, mittag_nonintrusive_2019, gamper_intrusive_2019, catellier_wenets_2019, lo_mosnet_2019, soni_novel_2016, patton_automos_2016, manocha_differentiable_2020}. A number of rule-based systems have the advantage of being perceptually-motivated and, to some extent, interpretable, but often present a narrow focus on specific types of signals or degradations, such as telephony signals or voice-over-IP (VoIP) degradations. Learning-based systems, on the other hand, are usually easy to repurpose to other tasks and degradations, but require considerable amounts of human annotated data. Both rule- and learning-based systems might additionally suffer from lack of generalization, and thus perform poorly on out-of-sample but still on-focus data.

Semi-supervised %(or self-supervised\footnote{These terms are often interchangeable. Here, we will adopt the semi-supervised term, as we always use the set of proposed techniques in conjunction with a purely supervised loss.}) 
learning is a possible strategy to deal with lack of annotations and poor generalization~\cite{chapelle_semisupervised_2006}. By leveraging both labeled and unlabeled data, 
%performance and model generalization can be improved (specially when annotations are scarce).
%, with possibly self-supervised tasks, 
it can bring substantial performance and generalization improvements, specially when annotations are scarce.
Semi-supervised learning is behind many recent advancements in machine 
\iflong
learning~\cite{van_engelen_survey_2020}, 
\else
learning,
\fi
and has been successfully applied to image quality assessment~\cite{ma_dipiq_2017, liu_exploiting_2019}. However, surprisingly, we find no purely semi-supervised approaches for audio or speech quality assessment. To the best of our knowledge, the few works that exploit unlabeled data for this task only make indirect use of it: they either exploit the output of other existing (rule-based) measures to complement the supervised loss~\cite{mittag_nonintrusive_2019}, or resort to pre-trained models or latent features~\cite{soni_novel_2016, patton_automos_2016, manocha_differentiable_2020}, usually coming from other/unrelated tasks.%\footnote{Interestingly, Manocha et al.~\cite{manocha_differentiable_2020} report some of their best accuracies when training their model from scratch, in a strictly supervised way, instead of using pre-training or fine-tuning.}. 

In this paper, we propose to learn a model of speech quality that combines multiple objectives, following a semi-supervised approach. We call it SESQA, for \underline{se}mi-supervised \underline{s}peech \underline{q}uality \underline{a}ssessment. SESQA learns from existing labeled data, together with limitless amounts of unlabeled or programmatically generated data, and produces speech quality scores, together with usable latent features and informative auxiliary outputs. Scores and outputs are concurrently optimized in a multi-task setting by 8~different but complementary objective criteria, with the idea that relevant cues are present in all of them. By flowing information through a shared latent space bottleneck, the considered objectives learn to cooperate, and promote better and more robust representations while discarding non-essential information~\cite{serra_towards_2018, pascual_learning_2019}. Additional design principles of SESQA include its lightweight and fast operation, its fully-differentiable nature, and its ability to deal with short-time raw audio frames at 48\,kHz (thus yielding a time-varying, dynamic estimate).

We evaluate SESQA against 9~existing approaches, under 3~different metrics, and using 3~different data sets. We focus on the reference-free setting~\cite{Loizou11BOOKCHAP}, but note that the proposed framework is easily extensible to learn from reference-based speech quality 
\iflong
scores (we provide a number of straightforward adaptations in Appendix~\ref{appendix:adapt}). 
\else
scores. 
\fi
On the considered metrics, we show that SESQA cuts the error of existing methods by more than 36\%, and that such a leap is due to the proposed semi-supervised approach using multiple criteria and auxiliary tasks. We also highlight the consistency of the speech quality estimates and the utility of the learned latent space. Finally, we show that SESQA also outperforms existing methods by more than 21\% in an out-of-sample post-hoc listening test, what suggests a good generalization to new recordings and listeners.

\section{The SESQA model}

\subsection{Optimization criteria and auxiliary tasks}
\label{sec:criteria}

A key driver of our work is to notice that additional evaluation criteria should be considered beyond correlation with mean opinion scores (MOS)~\cite{Loizou11BOOKCHAP} of speech quality. Importantly, we decide to also learn from such additional evaluation criteria. Another fundamental aspect of our work is to realize that there are further objectives, data sets, and tasks that can complement those criteria and help learning a more robust representation of speech quality and scores. The combination of multiple learning/evaluation criteria (Sec.~\ref{sec:criteria_eval}) and auxiliary tasks (Sec.~\ref{sec:criteria_additional}) is what gives birth to SESQA.

\subsubsection{Learning/evaluation criteria}
\label{sec:criteria_eval}

\minisection{Mean opinion score} The principal and almost unique criterion considered by existing approaches is the MOS error. In learning-based approaches, a supervised regression problem is usually set, such that
\begin{equation*}
L^{\text{MOS}} = \left\| s^\ast_i - s_i \right\| ,
\end{equation*}
where $s^\ast_i$ is the MOS ground truth, $s_i$ is the score predicted by the model, and $\|~\|$ corresponds to some norm. Throughout this work, we use the L1 norm (mean absolute error). 

SESQA predicts scores $s_i$ from a latent representation $\ve{z}_i$ by using a linear unit and a sigmoid activation $\sigma$: $s_i=1+4\sigma(\ve{w}^{\text{T}}\ve{z}_i+b)$, where coefficients 1 and 4 adapt the score to MOS values between 1 and 5. The latent representation $\ve{z}_i$ is obtained by encoding a raw audio frame $\ve{x}_i$ through a neural network encoder (we detail SESQA's architecture in Sec.~\ref{sec:architecture}).

\minisection{Pairwise ranking} Besides MOS, another intuitive but often overlooked notion in quality assessment is pairwise rankings~\cite{ma_dipiq_2017, liu_exploiting_2019}: if a speech signal $\ve{x}_j$ is a programmatically degraded version of the same (originally `clean', or `cleaner') utterance $\ve{x}_i$, then their scores should reflect such relation, that is, $s_i\ge s_j$. This notion can be introduced in a training schema by considering learning-to-rank strategies~\cite{burges_learning_2005}. In our case, we follow a margin loss formulation~\cite{liu_exploiting_2019}:
\begin{equation*}
L^{\text{RANK}} = \max\left( 0, s_j - s_i + \alpha \right) ,
\end{equation*}
where $\alpha=0.3$ is a margin constant. %This loss maximizes the difference between $s_i$ and $s_j$ up to a threshold $\alpha$, promoting lower values for $\ve{x}_i$ (the more degraded version) and higher values for $\ve{x}_j$ (the less degraded version).

In SESQA, we programmatically generate pairs $\{\ve{x}_i,\ve{x}_j\}$ by considering a number of data sets with clean speech and a pool of several degradation functions (we overview our data and methodology in Sec.~\ref{sec:methodology}). Additionally, we gather random pairs from annotated data, assigning indices $i$ and $j$ depending on the corresponding $s^\ast$, such that the element of the pair with a larger $s^\ast$ gets index $i$. For pairs coming from annotated data, we use  $\alpha'=\min(\alpha,s^\ast_i-s^\ast_j)$.

\minisection{Score consistency} Consistency is another overlooked notion in audio quality assessment~\cite{manjunath_limitations_2009}: if two signals $\ve{x}_k$ and $\ve{x}_l$ are extracted from the same source and differ by just a few audio samples, or if the difference between two signals $\ve{x}_k$ and $\ve{x}_l$ is perceptually irrelevant~\cite{kollmeier_perception_2008}, then their scores should be essentially the same, that is, $s_k=s_l$. Complementarily, if two signals $\ve{x}_i$ and $\ve{x}_j$ are perceptually distinguishable, then their score difference should be above a certain margin $\beta$, that is, $|s_i-s_j|\geq\beta$. Notice that these two notions can be extended to pairs of pairs, by considering the consistency between pairs of score differences. %In our current implementation, we only extend the first notion: if we have two signals $\ve{x}_{i_k}$ and $\ve{x}_{j_k}$ whose score difference is $s_{j_k}-s_{i_k}$ (with $\ve{x}_i$ having more degradation than $\ve{x}_j$), we can also enforce consistency in the difference of differences. That is, with $\ve{x}_{i_k}$ and $\ve{x}_{j_k}$ that are perceptually the same as $\ve{x}_{i_l}$ and $\ve{x}_{j_l}$, respectively, we should tend to $s_{j_k}-s_{i_k}=s_{j_l}-s_{i_l}$. 
In our current implementation, we only extend the first notion: if we have two signals $\ve{x}_{i_k}$ and $\ve{x}_{j_k}$ that are respectively perceptually the same as $\ve{x}_{i_l}$ and $\ve{x}_{j_l}$ (with $\ve{x}_j$ having more degradation than $\ve{x}_i$, signals $k$ and $l$ extracted from those), score differences should tend to be equal, that is, $s_{i_k}-s_{j_k}=s_{i_l}-s_{j_l}$. 

Taking all three notions into account, we propose the consistency loss
\begin{equation*}
\begin{split}
L^{\text{CONS}} & = \frac{1}{4}\left( | s_k-s_l | + | s_{i_k}-s_{j_k}-(s_{i_l}-s_{j_l}) | \right) + \\ 
                & + \frac{1}{2\beta}\left( 1- \min\left(|s_i-s_j|,\beta\right) \right),
\end{split}
\end{equation*}
where $\beta=0.1$ is another margin constant. %We see there are three main terms in this loss function: the first one promotes small differences between $s_k$ and $s_l$, the second one promotes small differences between different score differences $s_{j_k}-s_{i_k}$ and $s_{j_l}-s_{i_l}$, and the third one promotes large differences between $s_i$ and $s_j$, up to some constant $\beta$. Coefficients $1/4$ and $1/2$ distribute weights equally across terms. 
We programmatically generate quadruples $\{\ve{x}_{i_k},\ve{x}_{i_l},\ve{x}_{j_k},\ve{x}_{j_l}\}$ by extracting them from pairs $\ve{x}_i$ and $\ve{x}_j$ using a random small delay below 100\,ms (we reuse the pairs $\{\ve{x}_i,\ve{x}_j\}$ from $L^{\text{RANK}}$). In addition to those, we consider pairs $\{\ve{x}_i,\ve{x}_j\}$ and $\{\ve{x}_k,\ve{x}_l\}$ from an existing just-noticeable difference (JND) data set~\cite{manocha_differentiable_2020}.% (data set details are summarized in Sec.~\ref{sec:methodology}).

\subsubsection{Auxiliary tasks}
\label{sec:criteria_additional}

\minisection{Same/different condition} With the data that we programmatically generate for $L^{\text{CONS}}$, we also have information on pairs of signals that correspond to the same degradation condition, that is, signals that have undergone the same degradation type and strength. We can include this information by considering the classification loss
\begin{equation*}
L^{\text{SD}} = \text{BCE}\left( \delta^{\text{SD}} , H^{\text{SD}}(\ve{z}_u,\ve{z}_v ) \right) ,
\end{equation*}
where BCE stands for binary cross-entropy, $\delta^{\text{SD}}\in\{0,1\}$ indicates if latent vectors $\ve{z}_u$ and $\ve{z}_v$ correspond to the same condition ($\{u,v\}\widehat{=}\{k,l\}$) or not ($\{u,v\}\widehat{=}\{i,j\}$), and $H$ is a small neural network that takes the concatenation of the two vectors and produces a single probability value (further details on $H$ are available in Sec.~\ref{sec:architecture}).

\minisection{Just-noticeable difference} If, as mentioned, we have access to pairs of signals with human annotations regarding their perceptual difference, we can reinforce this notion in our latent space with another classification loss
\begin{equation*}
L^{\text{JND}} = \text{BCE}\left( \delta^{\text{JND}} , H^{\text{JND}}(\ve{z}_u,\ve{z}_v ) \right) ,
\end{equation*}
where $\delta^{\text{JND}}\in\{0,1\}$ indicates if latent representations $\ve{z}_u$ and $\ve{z}_v$ correspond to a JND or not~\cite{manocha_differentiable_2020}.

\minisection{Degradation type} Another advantage of programmatically generated data is that, if we start from signals that we consider clean or without noticeable degradations, we know which degradations have been applied. With that, we build a multi-class classification loss
\begin{equation*}
L^{\text{DT}} = \sum_n \text{BCE}\left( \delta^{\text{DT}}_n , H^{\text{DT}}_n(\ve{z}_i) \right) ,
\end{equation*}
where $\delta^{\text{DT}}_n\in\{0,1\}$ indicates whether the latent representation $\ve{z}_i$ contains degradation $n$ or not (degradations are summarized in Sec.~\ref{sec:methodology}). We also include the case where there is no degradation as one of the $n$ possibilities, therefore constituting on its own a binary clean/degraded classifier. 

\minisection{Degradation strength} At the moment of applying a degradation to a signal, we usually have to decide a degradation strength. Therefore, we can add the corresponding regressors
\begin{equation*}
L^{\text{DS}} = \sum_n \left\| \zeta^{\text{DS}}_n - H^{\text{DS}}_n(\ve{z}_i) \right\| ,
\end{equation*}
where $\zeta^{\text{DS}}_n\in[0,1]$ indicates the strength of degradation $n$. 

\minisection{Other quality assessment measures} Finally, since we generate pairs $\{\ve{x}_i,\ve{x}_j\}$, we can always compute existing reference-based (or reference-free) quality measures over those pairs and learn from them. We do that with a pool of regression losses
\begin{equation*}
L^{\text{MR}} = \sum_m \left\| \zeta^{\text{MR}}_m - H^{\text{MR}}_m(\ve{z}_i,\ve{z}_j) \right\| ,
\end{equation*}
where $\zeta^{\text{MR}}_m\in\mathbb{R} $ is the value for measure $m$ computed on $\{\ve{x}_i,\ve{x}_j\}$ (we normalize $\zeta^{\text{MR}}_m$ to have zero mean and unit variance based on training data). In this work, we consider reference-based measures PESQ, CSIG, CBAK, COVL, SSNR, LLR, WSSD, STOI, SISDR, Mel cepstral distortion, and log-Mel-band 
\iflong
distortion~\cite{Loizou11BOOKCHAP, Rix01ICASSP, taal_algorithm_2011, roux_sdr_2019}.
\else
distortion~\cite{Loizou11BOOKCHAP, Rix01ICASSP, taal_algorithm_2011}.
\fi

\subsection{Architecture}
\label{sec:architecture}

\begin{figure}[t]
  \hfill
  \includegraphics[width=1\linewidth]{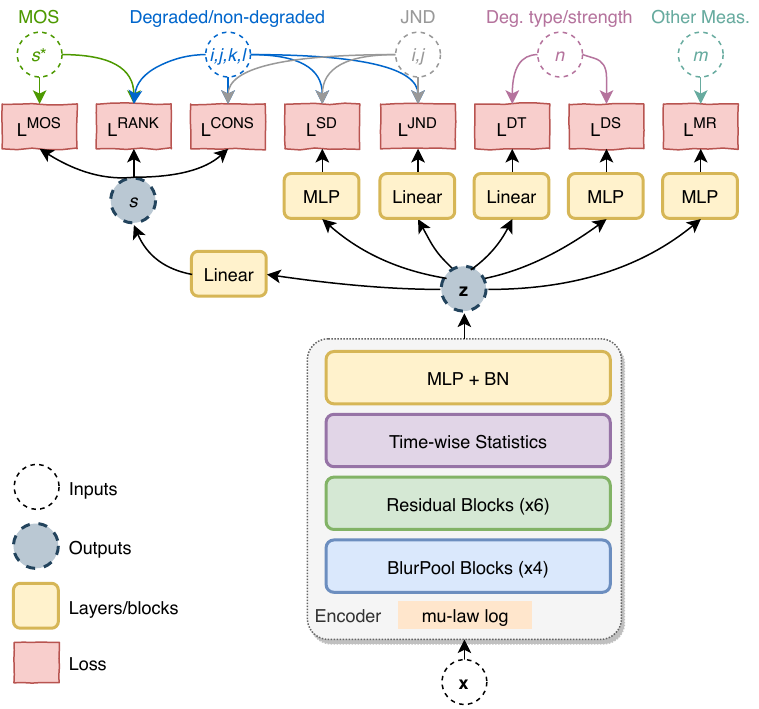}
  \caption{Block diagram of the SESQA model.}
  \label{fig:blocks}
\end{figure}

SESQA is composed of an encoder and a series of heads $H$ (Fig.~\ref{fig:blocks}). The encoder takes raw audio frames $\ve{x}$ and maps them to latent space vectors $\ve{z}$. Heads $H$ take these latent vectors $\ve{z}$ and compute the outputs for the considered criteria and tasks (Sec.~\ref{sec:criteria}). As mentioned, a linear unit with sigmoid activation is used to project $\ve{z}$ to $s$. When dealing with pairs $\{\ve{z}_i,\ve{z}_j\}$, $H$ takes their concatenation as input.

The encoder consists of 4~main stages. Firstly, we transform the distribution of $\ve{x}$ by applying a $\mu$-law companding (no quantization) with a learnable $\mu$, which we initialize to 8. Next, we employ 4~pooling blocks, each consisting of convolution, batch normalization (BN), rectified linear unit (ReLU) activation, and BlurPool~\cite{zhang_making_2019}. We use 32, 64, 128, and 256 filters with a kernel width of 4 and a downsampling factor of 4. Next, we employ 6~residual blocks formed by a BN pre-activation and 3~blocks of ReLU, convolution, and BN. We use 512, 512, and 256 filters with kernel widths 1, 3, and 1, respectively, and add the residual connection by parametric linear averaging: $\ve{h}'=\ve{a}' \ve{h} + (1-\ve{a}')F(\ve{h})$, where $\ve{a}'=\sigma(\ve{a})$ is a vector of learnable parameters between 0 and 1 and $F$ is the residual network (we initialize all components of $\ve{a}$ to 3 so that training starts with mostly a bypass from $\ve{h}$ to $\ve{h}'$). After the residual blocks, we compute time-wise statistics, taking the per-channel mean and standard deviation. This step aggregates all temporal information into a single vector of 2$\times$256~dimensions. We perform BN on such vector and input it to a multi-layer perceptron (MLP) formed by two linear layers with BN, using a ReLU activation in the middle. We employ 1024 and 200~units.

The multiple heads $H$ consist of either linear layers or two-layer MLPs with 400~units, all with BN at the end. We prefer simple heads in order to encourage the encoder, and not the heads, to learn high-level features that can be successfully exploited even by networks with limited capacity~\cite{serra_towards_2018, pascual_learning_2019}. The decision of whether to use a linear layer or an MLP is based on the idea that the more relevant the auxiliary task, the less capacity should the head have.
%but with the idea that, intuitively, more relevant criteria should be allocated less capacity. 
This way, we empirically choose a linear layer for the score $s$ and the JND and DT heads. 
%Notice that setting linear layers for these three heads will confer interesting properties to the latent space. For example, promoted by $s$ and $L^{\text{JND}}$ losses, we expect \ve{z} to reflect coherent `distances' between latent representations. But we also expect audio with similar degradation types to group/clusters together, due to the $L^{\text{DT}}$ linear head. An informal validation of these intuitions is provided in Appendix~\ref{appendix:results}.
Notice that setting linear layers for these three heads will provide interesting properties to the latent space, making it reflect `distances' between latent representations, due to $s$ and $L^{\text{JND}}$, and promoting groups/clusters of degradation types, due to $L^{\text{DT}}$ 
\iflong
(a validation of these intuitions is provided in Appendix~\ref{appendix:results}).
\else
(Sec.~\ref{sec:results}).
\fi

%\subsection{Further details}
%\label{sec:details}

We train SESQA with the RangerQH optimizer~\cite{wright_new_2019}, using default parameters and a learning rate of $10^{-3}$. We decay the learning rate by a factor of $1/5$ at 70 and 90\% of training. To favor generalization and slightly improve performance, we also employ stochastic weight averaging~\cite{izmailov_averaging_2018} during the last training epoch. Since after a few iterations all losses are within a similar scale, we do not perform any loss weighting. %We observe that all validation losses first go down for a number of steps and that, after some point, remain stable or with negligible fluctuations.

%%%%%%%%%%%%%%%%%%%%%%%%%%%%%%%%%%%%%%%%%%%%%%%%%%%%%%%%%%%%%%%%%%%%%%%%%%%%%%%%%%%%%

\section{Data and methodology}
\label{sec:methodology}

We use 3~MOS data sets, two internal and a publicly-available one. The first internal data set consists of 1,109~recordings and a total of 1.5\,h of audio, featuring mostly user-generated content (UGC). The second internal dataset consists of 8,016~recordings and 15\,h of audio, featuring telephony and VoIP degradations. The third data set is TCD-VoIP~\cite{harte_tcdvoip_2015}, which consists of 384~recordings and 0.7\,h of audio, featuring a number of VoIP degradations. Another data set that we use is the JND data set~\cite{manocha_differentiable_2020}, which consists of 20,797~pairs of recordings and 28\,h of audio. 
\iflong
More details can be found in Appendix~\ref{appendix:data}. 
\fi
For the programmatic generation of data, we use a pool of internal and public data sets, and we generate 70,000~quadruples conforming 78\,h audio. We employ a total of 37~possible degradations, including additive background noise, hum noise, clipping, sound effects, packet losses, phase distortions, and a number of audio codecs. 
\iflong
More details can be found in Appendix~\ref{appendix:degradations}.
\else
More details on annotated and programmatically generated data can be found in~\cite{serra_sesqa_2020}.
\fi

We compare SESQA with ITU-P563~\cite{Malfait11TASL}, two approaches based on feature losses, one using JND~\cite{manocha_differentiable_2020} (FL-JND) and another one using PASE~\cite{pascual_learning_2019} (FL-PASE), SRMR~\cite{Falk10TASLP}, AutoMOS~\cite{patton_automos_2016}, Quality-Net~\cite{fu_qualitynet_2018}, WEnets~\cite{catellier_wenets_2019}, CNN-ELM~\cite{gamper_intrusive_2019}, and NISQA~\cite{mittag_nonintrusive_2019}. We re-implement some of them to fit our training and evaluation pipelines and adapt them to work at 48\,kHz, if needed/possible. We note that FL, AutoMOS, and NISQA make use of partial additional data beyond MOS, thus being weakly semi-supervised approaches (we discussed the `weak' terminology in Sec.~\ref{sec:intro}). 
\iflong
More details on baseline approaches can be found in Appendix~\ref{appendix:approaches}.
\fi

We put all approaches under the same setting, choosing their best optimizers and hyper-parameters on the validation set. We train with weakly-labeled frames of 1\,s for 5~epochs, performing data augmentation and reusing MOS data inside an epoch (we define an epoch as a full pass over the programmatically generated data). We use random scaling, phase inversion, and temporal sampling as data augmentation. For evaluation, we employ $L^{\text{MOS}}$ and $L^{\text{CONS}}$, and compute the ratio of incorrectly classified rankings $R^{\text{RANK}}$~\cite{ma_dipiq_2017, liu_exploiting_2019} (we report $R^{\text{RANK}}$ instead of $L^{\text{RANK}}$ for interpretability). In addition, we compute a summary error $E^{\text{TOTAL}}=0.5L^{\text{MOS}}+R^{\text{RANK}}+L^{\text{CONS}}$ (we introduce the 0.5~weight to compensate for the different range). We perform 5-fold cross-validation and report average errors.

%%%%%%%%%%%%%%%%%%%%%%%%%%%%%%%%%%%%%%%%%%%%%%%%%%%%%%%%%%%%%%%%%%%%%%%%%%%%%%%%%%%%%

\section{Results}
\label{sec:results}

We start by comparing SESQA with existing approaches 
\iflong
(Table~\ref{tab:avg_error}). 
\else
(Table~\ref{tab:avg_error}; results for individual data sets and further experiments can be found at~\cite{serra_sesqa_2020}).
\fi
Overall, we observe that all approaches clearly outperform the random baseline, and that around half of them achieve an error comparable to the variability between human scores ($L^{\text{MOS}}$ estimated by taking the standard deviation across listeners and averaging across utterances). We also observe that many of the existing approaches report decent consistencies, with $L^{\text{CONS}}$ around~0.1, six times lower than the random baseline. However, noticeably, existing approaches yield considerable errors for relative pairwise rankings ($R^{\text{RANK}}$). SESQA outperforms all existing approaches in all considered evaluation metrics by a large margin, including the standard $L^{\text{MOS}}$. The only exception to the previous statement is with the $L^{\text{CONS}}$ metric of the ITU-P563 approach, which nonetheless has a high $L^{\text{MOS}}$ and an almost random $R^{\text{RANK}}$. With the summary metric $E^{\text{TOTAL}}$, SESQA cuts the error of the best existing approach by 36\%. 

\begin{table}[t]
  \centering
  \begin{tabular}{p{2.5cm}cccc}
    \hline
    \textbf{Approach} & \textbf{$L^{\text{MOS}}$} & \textbf{$R^{\text{RANK}}$} & \textbf{$L^{\text{CONS}}$} & \textbf{$E^{\text{TOTAL}}$} \\
    \hline
    Human           & 0.679 & n/a   & n/a   & n/a   \\
    Random score    & 1.219 & 0.500 & 0.614 & 1.724 \\
    \hline
    ITU-P563        & 0.982 & 0.498 & 0.050 & 1.042 \\
    FL-JND          & 0.899 & 0.365 & 0.093 & 0.908 \\
    SRMR            & 0.854 & 0.351 & 0.071 & 0.849 \\
    FL-PASE         & 0.735 & 0.324 & 0.105 & 0.796 \\
    AutoMOS         & 0.537 & 0.311 & 0.212 & 0.792 \\
    Quality-Net     & 0.657 & 0.349 & 0.087 & 0.765 \\
    WEnet           & 0.660 & 0.258 & 0.125 & 0.713 \\
    NISQA           & 0.556 & 0.243 & 0.123 & 0.644 \\
    CNN-ELM         & 0.511 & 0.220 & 0.145 & 0.621 \\
    \hline
    SESQA (ours)    & 0.474 & 0.090 & 0.067 & 0.394 \\
    \hline
  \end{tabular}
  \caption{Comparison with existing approaches. Average errors across the three considered data 
  \iflong
  sets (results for individual data sets are available in Appendix~\ref{appendix:results}).
  \else
  sets.
  \fi
  }
  \label{tab:avg_error}
\end{table}

We now study the effect that the considered criteria/tasks have on the performance of SESQA (Table~\ref{tab:avg_ablation}). First of all, we observe that errors never decrease by removing a single criterion. This indicates that none of them is harmful in terms of performance. Next, we observe that there are some relevant criteria that, if removed, have a considerable impact (for example $L^{\text{MOS}}$ and $L^{\text{RANK}}$). However, the absence of one of such relevant criteria does not yet produce the average error of existing approaches ($E^{\text{TOTAL}}$, Table~\ref{tab:avg_error}). Regarding less relevant tasks, we should note that we still find them useful for the outputs that they produce (for example, knowing if a pair of signals present a JND difference) or for the properties they confer to the organization of the latent space $\ve{z}$ (examples of latent space distances and degradation type clustering are shown 
\iflong
in Appendix~\ref{appendix:results}). 
\else
in~\cite{serra_sesqa_2020}). 
\fi
Finally, it is also interesting to highlight that considering the $L^{\text{MOS}}$ criterion alone (last row, Table~\ref{tab:avg_ablation}) yields a performance that is on par with the best-performing existing approaches (NISQA and CNN-ELM, Table~\ref{tab:avg_error}). Overall, this demonstrates that considering multiple optimization criteria and tasks is key for achieving outstanding performance, and empirically justifies a semi-supervised approach to audio quality assessment like SESQA.

\begin{table}[t]
  \centering
  \begin{tabular}{p{2.5cm}cccc}
    \hline
    \textbf{Approach} & \textbf{$L^{\text{MOS}}$} & \textbf{$R^{\text{RANK}}$} & \textbf{$L^{\text{CONS}}$} & \textbf{$E^{\text{TOTAL}}$} \\
    \hline
    SESQA                       & 0.474 & 0.090 & 0.067 & 0.394 \\
    Without $L^{\text{MOS}}$    & 0.839 & 0.079 & 0.044 & 0.543 \\
    Without $L^{\text{RANK}}$   & 0.492 & 0.201 & 0.061 & 0.508 \\
    Without $L^{\text{CONS}}$   & 0.441 & 0.096 & 0.130 & 0.447 \\
    Without $L^{\text{SD}}$     & 0.482 & 0.091 & 0.067 & 0.399 \\
    Without $L^{\text{JND}}$    & 0.475 & 0.089 & 0.067 & 0.394 \\
    Without $L^{\text{DT}}$     & 0.476 & 0.089 & 0.067 & 0.394 \\
    Without $L^{\text{DS}}$     & 0.479 & 0.090 & 0.067 & 0.396 \\
    Without $L^{\text{MR}}$     & 0.488 & 0.093 & 0.066 & 0.403 \\
    Only $L^{\text{MOS}}$       & 0.480 & 0.265 & 0.137 & 0.643 \\
    \hline
  \end{tabular}
  \caption{Loss ablation study. Average errors across the three considered data \iflong
  sets (results for individual data sets are available in Appendix~\ref{appendix:results}).
  \else
  sets.
  \fi
  }
  \label{tab:avg_ablation}
\end{table}

To further assess the generalization capabilities of the considered approaches, we also perform a post-hoc informal test with out-of-sample data. For that, we choose 20~new recordings from UGC, featuring clean or production-quality speech, and speech with degradations such as real background noise, codec artifacts, or microphone distortion. We then ask a new set of listeners to rate the quality of the 
\iflong
recordings with a score between 1 and 5, 
\else
recordings, 
\fi
and compare their ratings with the ones produced by models pre-trained on our internal UGC data set (Table~\ref{tab:posthoc_test}). We see that the ranking of existing approaches changes, showing that some are better than others at generalizing to out-of-sample data. Nonetheless, SESQA still outperforms them in all metrics and by a large margin. Noticeably, it cuts the $L^{\text{MOS}}$ of the best existing approach by 21\%, which is much more than the relative $L^{\text{MOS}}$ difference 
\iflong
observed for in-sample data, which was 7\% (from Table~\ref{tab:avg_error}). 
\else
of 7\% for in-sample data (Table~\ref{tab:avg_error}).
\fi
This indicates that SESQA generalizes better to out-of-sample but related data. %, and it is the only method to report a Pearson's correlation coefficient equal to the average correlation of individual listeners to the mean score.

\begin{table}[t]
  \centering
  \begin{tabular}{p{2.5cm}ccc}
    \hline
    \textbf{Approach} & \textbf{$L^{\text{MOS}}$ (Std)} & \textbf{$\rho_{\text{P}}$} & \textbf{$\rho_{\text{S}}$} \\
    \hline
    Human           & 0.62 \,(n/a)\,\,  & 0.87  & 0.86   \\
    Random score    & 1.18 (0.87) & 0.15  & 0.16  \\
    \hline
    FL-JND          & 1.19 (0.80) & 0.01 & 0.30 \\
    ITU-P563        & 0.92 (0.50) & 0.05 & 0.04 \\
    SRMR            & 0.83 (0.78) & 0.25 & 0.31 \\
    NISQA           & 0.74 (0.68) & 0.61 & 0.60 \\
    FL-PASE         & 0.64 (0.55) & 0.63 & 0.67 \\
    Quality-Net     & 0.64 (0.50) & 0.82 & 0.78 \\
    WEnet           & 0.55 (0.44) & 0.74 & 0.72 \\
    AutoMOS         & 0.48 (0.45) & 0.82 & 0.82 \\
    CNN-ELM         & 0.47 (0.37) & 0.84 & 0.81 \\
    \hline
    SESQA (ours)    & 0.37 (0.34) & 0.87 & 0.89 \\
    \hline
  \end{tabular}
  \caption{Errors and correlations for the post-hoc listening test. To complement $L^{\text{MOS}}$, we here additionally consider Pearson's ($\rho_{\text{P}}$) and Spearman's ($\rho_{\text{S}}$) correlation.}
  \label{tab:posthoc_test}
\end{table}

%%%%%%%%%%%%%%%%%%%%%%%%%%%%%%%%%%%%%%%%%%%%%%%%%%%%%%%%%%%%%%%%%%%%%%%%%%%%%%%%%%%%%

\section{Discussion}

Given the difficulties of collecting listener scores and the potential lack of generalization of existing approaches, we believe that semi-supervised learning represents the current best way forward in automatic audio or speech quality assessment. In particular, we are encouraged by the capacity of semi-supervised models for leveraging virtually infinite amounts of data, and for combining those with multiple optimization criteria and complementary auxiliary tasks. %, including the regression of standard MOS and the output of other existing speech quality measures. 
Since a single score may be insufficient 
%for a number of applications, we think that the auxiliary outputs by SESQA's heads (including, e.g., degradation type or strength) can yield to \santi{no sobra el "to"?} new perspectives for audio quality analytics.
in a number of final applications, we are also enthusiastic for the reuse of auxiliary outputs and learnt latents in such applications or further downstream tasks.

%%%%%%%%%%%%%%%%%%%%%%%%%%%%%%%%%%%%%%%%%%%%%%%%%%%%%%%%%%%%%%%%%%%%%%%%%%%%%%%%%%%%%

%\section{Acknowledgements}

%\joan{Maybe we do not need...}

%\vfill\pagebreak

\bibliographystyle{IEEEbib}
\bibliography{sesqa_biblio}

\iflong
\clearpage
\appendix
\onecolumn

\begin{center}
\textbf{\large{APPENDIX}}
\end{center}
\vspace{2.5mm}
\setstretch{1.1}

%%%%%%%%%%%%%%%%%%%%%%%%%%%%%%%%%%%%%%%%%%%%%%%%%%%%%%%%%%%%%%%%%%%%%%%%%%%%%%%%%%%%%

\section{Computing scores with a reference signal}
\label{appendix:adapt}

If we want to compute scores $s$ in a reference-based setting instead of a reference-free one, we just need to pass the two signals $\ve{x}_i$ and $\ve{x}_j$ through the encoder and obtain the corresponding latents $\ve{z}_i$ and $\ve{z}_j$. Then, for instance, we can compute $s_{ij}=1+4\sigma(\ve{w}^{\text{T}}\ve{z}_i-\ve{w}^{\text{T}}\ve{z}_j+b)$, using a linear unit for both latents. Other options are to compute a single score from a latent vector difference, $s_{ij}=1+4\sigma(\ve{w}
^{\text{T}}(\ve{z}_i-\ve{z}_j)+b)$, or to concatenate latents and use a layer that is double the size, $s_{ij}=1+4\sigma(\ve{w}^{\text{T}}[\ve{z}
^{\text{T}}_i;\ve{z}^{\text{T}}_j]^{\text{T}}+b)$. Additional perspectives include replacing vector differences or linear layers by more complicated nonlinear, parametric, and/or learnable functions.

%%%%%%%%%%%%%%%%%%%%%%%%%%%%%%%%%%%%%%%%%%%%%%%%%%%%%%%%%%%%%%%%%%%%%%%%%%%%%%%%%%%%%

\section{Data}
\label{appendix:data}

As mentioned, in our semi-supervised approach we employ 3~types of data: MOS data, JND data, and programmatically generated data. The additional out-of-sample data set used in the post-hoc listening test is summarized in the main paper, and its degradation characteristics resemble the ones in our internal UGC data set (see below).

\subsection{MOS data}

We train and evaluate on 3~different MOS data sets of different size and characteristics:

\begin{enumerate}

\item Internal UGC data set --- This data set consists of 1,109~recordings of UGC, adding up to a total of 1.5\,h of audio. All recordings are converted to mono WAV PCM at 48\,kHz and normalized to have the same loudness. Utterances range from single words to few sentences, uttered by both male and female speakers in a variety of conditions, using different languages (mostly English, but also Chinese, Russian, Spanish, etc.). Common degradations in the recordings include background noise (street, cafeteria, wind, background TV/radio, other people's speech, etc.), reverb, bandwidth reduction (low-pass down to 3\,kHz), and coding artifacts (MP3, OGG, AAC, etc.). Quality ratings were collected with the help of a pool of 10~expert listeners with at least a few years of experience in audio processing/engineering. Recordings have between 4 and 10 ratings, which were obtained by following standard procedures like the ones described by IEEE and ITU (see~\cite{Loizou11BOOKCHAP} and references therein).

\item Internal telephony/VoIP data set --- This data set consists of 8,016~recordings with typical telephony and VoIP degradations, adding up to a total of 15\,h of audio. Besides a small percentage, all audios are originally recorded at 48\,kHz before further processing and normalized to have the same loudness. Recordings contains two sentences separated by silence and have a duration between 5 and 15\,s, following a protocol similar to ITU-P800. Male and female utterances are balanced and different languages are present (English, French, Italian, Czech, etc.). Common degradations include packet losses (between 20 and 60\,ms), bandwidth reduction (low-pass down to 3\,kHz), additive synthetic noise (different SNRs), and coding artifacts (G772, OPUS, AC3, etc.). Quality ratings are provided by a pool of regular listeners, with each recording having between 10 and 15~ratings. Ratings were obtained by following the standard procedure described by ITU (see~\cite{Loizou11BOOKCHAP} and references therein).

\item TCD-VoIP data set --- This is a public dataset available online at \url{http://www.mee.tcd.ie/~sigmedia/Resources/TCD-VoIP}. It consists of 384~recordings with common VoIP degradations, adding up to a total of 0.7\,h. A good description of the data set is provided in the original reference~\cite{harte_tcdvoip_2015}. Despite also being VoIP degradations, a number of them differ from our internal telephony/VoIP data set (both in type and strength).

\end{enumerate}

\subsection{JND data}

We also use JND data for training. We resort to the data set compiled by Manocha et al.~\cite{manocha_differentiable_2020}, which is available at \url{https://github.com/pranaymanocha/PerceptualAudio}. The data set consists of 20,797~pairs of ``perturbed'' recordings (28\,h of audio), each pair coming from the same utterance, with annotations of whether such perturbations are pairwise noticeable or not. Annotations were crowd-sourced from Amazon Mechanical Turk following a specific procedure~\cite{manocha_differentiable_2020}. Perturbations correspond to additive linear background noise, reverb, and coding/compression.

\subsection{Programmatically generated data}

We compute quadruples $\{\ve{x}_{i_k},\ve{x}_{i_l},\ve{x}_{j_k},\ve{x}_{j_l}\}$ of programmatically generated data. To do so, we start from a list of 10~data sets of audio at 48\,kHz that we consider clean and without processing. This includes private/proprietary data sets, and public data sets such as VCTK~\cite{yamagishi_cstr_2019}, RAVDESS~\cite{livingstone_ryerson_2018}, or TSP Speech (\url{http://www-mmsp.ece.mcgill.ca/Documents/Data/}). For the experiments reported in this paper, we use 50,000~quadruples for training, 10,000 for validation, and 10,000 for testing. To form every quadruple, we proceed as follows:
\begin{itemize}
\item Uniformly sample a data set and uniformly sample a file from it.
\item Uniformly sample a 1.1\,s frame, avoiding silent or majorly silent frames. Normalize it to have a maximum absolute amplitude of 1.
\item With probabilities 0.84, 0.12, and 0.04 sample zero, one, or two degradations from the pool of available degradations (see below). If zero degradations, the signal directly becomes $\ve{x}_i$. Otherwise, we uniformly choose a strength for each degradation and apply them sequentially to generate $\ve{x}_i$.
\item With probabilities 0.75, 0.2, 0.04, and 0.01 sample one, two, three, or four degradations from the pool of available degradations (see below). Uniformly select strengths and apply them to $\ve{x}_i$ sequentially to generate $\ve{x}_j$. 
\item Uniformly sample a time delay between 0 and 100\,ms. Extract 1\,s frames $\ve{x}_{i_k}$ and $\ve{x}_{i_l}$ from $\ve{x}_i$ using such delay, and do the same for $\ve{x}_{j_k}$ and $\ve{x}_{j_l}$ from $\ve{x}_j$.
\item Store $\{\ve{x}_{i_k},\ve{x}_{i_l},\ve{x}_{j_k},\ve{x}_{j_l}\}$, together with the information of degradation type and strength.
\end{itemize}
In total, we use 78\,h of audio: $1\times4\times(50000+10000+10000)/3600=77.77\,\text{h}$.

%%%%%%%%%%%%%%%%%%%%%%%%%%%%%%%%%%%%%%%%%%%%%%%%%%%%%%%%%%%%%%%%%%%%%%%%%%%%%%%%%%%%%

\section{Degradations and strengths}
\label{appendix:degradations}

We consider 37~possible degradations with their corresponding strengths. Strengths have been set such that, at their minimum, they were perceptually noticeable by the authors. Note that, in some cases, the strengths chosen below are only one aspect of the whole degradation and that, for other relevant aspects, we randomly sample between empirically chosen values. For instance, for the case of the reverb effect, we select the SNR as the main strength, but we also randomly choose a type of reverb, a width, a delay, etc.
\begin{enumerate}
\item Additive noise --- With probability 0.29, sample a noise frame from the available pool of noise data sets. Add it to $\ve{x}$ with an SNR between 35 and $-$15\,dB. Noise data sets include private/proprietary data sets and public data sets such as ESC~\cite{piczak_esc_2015} or FSDNoisy18k~\cite{fonseca_learning_2019}. This degradation can be applied to the whole frame or, with probability 0.25, to just part of it (minimum 300\,ms).
\item Colored noise --- With probability 0.07, generate a colored noise frame with uniform exponent between 0 and 0.7. Add it to $\ve{x}$ with an SNR between 45 and $-$15\,dB. This degradation can be applied to the whole frame or, with probability 0.25, to just part of it (minimum 300\,ms).
\item Hum noise --- With probability 0.035, add tones around 50 or 60\,Hz (sine, sawtooth, square) with an SNR between 35 and $-$15\,dB. This degradation can be applied to the whole frame or, with probability 0.25, to just part of it (minimum 300\,ms).
\item Tonal noise --- With probability 0.011, same as before but with frequencies between 20 and 12,000\,Hz.
\item Resampling --- With probability 0.011, resample the signal to a frequency between 2 and 32\,kHz and convert it back to 48\,kHz.
\item $\mu$-law quantization --- With probability 0.011, apply $\mu$-law quantization between 2 and 10\,bits.
\item Clipping --- With probability 0.011, clip between 0.5 and 99\% of the signal.
\item Audio reverse --- With probability 0.05, temporally reverse the signal.
\item Insert silence --- With probability 0.011, insert between 1 and 10 silent sections of lengths between 20 and 120\,ms.
\item Insert noise --- With probability 0.011, same as above but with white noise.
\item Insert attenuation --- With probability 0.011, same as above but attenuating the section by multiplying by a maximum linear gain of 0.8.
\item Perturb amplitude --- With probability 0.011, same as above but inserting multiplicative Gaussian noise.
\item Sample duplicate --- With probability 0.011, same as above but replicating previous samples.
\item Delay --- With probability 0.035, add a delayed version of the signal (single- and multi-tap) using a maximum of 500\,ms delay.
\item Extreme equalization --- With probability 0.006, apply an equalization filter with a random Q and a gain above 20\,dB or below $-$20\,dB.
\item Band-pass --- With probability 0.006, apply a band-pass filter with a random Q at a random frequency between 100 and 4,000\,Hz.
\item Band-reject --- With probability 0.006, same as above but rejecting the band.
\item High-pass --- With probability 0.011, apply a high-pass filter at a random cutoff frequency between 150 and 4,000\,Hz.
\item Low-pass --- With probability 0.011, apply a low-pass filter at a random cutoff frequency between 250 and 8,000\,Hz.
\item Chorus --- With probability 0.011, add a chorus effect with a linear gain between 0.15 and 1.
\item Overdrive --- With probability 0.011, add an overdrive effect with a gain between 12 and 50\,dB.
\item Phaser --- With probability 0.011, add a phaser effect with a linear gain between 0.1 and 1.
\item Reverb --- With probability 0.035, add reverberation with an SNR between $-$5 and 10\,dB.
\item Tremolo --- With probability 0.011, add a tremolo effect with a depth between 30 and 100\%.
\item Griffin-Lim reconstruction --- With probability 0.023, perform a Griffin-Lim reconstruction of an STFT of the signal. The STFT is computed using random window lengths and 50\% overlap.
\item Phase randomization --- With probability 0.011, same as above but with random phase information.
\item Phase shuffle --- With probability 0.011, same as above but shuffling window phases in time.
\item Spectrogram convolution --- With probability 0.011, convolve the STFT of the signal with a 2D kernel. The STFT is computed using random window lengths and 50\% overlap.
\item Spectrogram holes --- With probability 0.011, apply dropout to the spectral magnitude with probability between 0.15 and 0.98.
\item Spectrogram noise --- With probability 0.011, same as above but replacing 0s by random values.
\item Transcoding MP3 --- With probability 0.023, encode to MP3 and back, using \texttt{libmp3lame} and between 2 and 96\,kbps (all codecs come from \texttt{ffmpeg}).
\item Transcoding AC3 --- With probability 0.035, encode to AC3 and back using between 2 and 96\,kbps.
\item Transcoding EAC3 --- With probability 0.023, encode to EAC3 and back using between 16 and 96\,kbps.
\item Transcoding MP2 --- With probability 0.023, encode to MP2 and back using between 32 and 96\,kbps.
\item Transcoding WMA --- With probability 0.023, encode to WMA and back using between 32 and 128\,kbps.
\item Transcoding OGG --- With probability 0.023, encode to OGG and back, using \texttt{libvorbis} and between 32 and 64\,kbps.
\item Transcoding OPUS --- With probability 0.046, encode to OPUS and back, using \texttt{libopus} and between 2 and 64\,kbps.
\end{enumerate}

%%%%%%%%%%%%%%%%%%%%%%%%%%%%%%%%%%%%%%%%%%%%%%%%%%%%%%%%%%%%%%%%%%%%%%%%%%%%%%%%%%%%%

\section{Considered Approaches}
\label{appendix:approaches}

We compare SESQA to 9~existing approaches:
\begin{enumerate}
\item ITU-P563~\cite{Malfait11TASL} --- This is a reference-free standard designed for narrowband telephony. We chose it because it was the best match for a reference-free standard that we had access to. We directly use the produced scores.
\item FL-JND --- Inspired by Manocha et al.~\cite{manocha_differentiable_2020}, we implement their proposed encoder architecture and train it on the JND task. Next, for each data set, we train a small MLP with a sigmoid output that takes latent features from all encoder layers as input and predicts quality scores.
\item FL-PASE --- We also take a PASE encoder~\cite{pascual_learning_2019} and train it with the tasks of JND, DT, and speaker identification. Next, for each data set, we train a small MLP with a sigmoid output that takes latent features from the last layer as input and predicts quality scores.
\item SRMR~\cite{Falk10TASLP} --- We use the measure from \url{https://github.com/jfsantos/SRMRpy} and employ a small MLP with a sigmoid output to adapt it to the corresponding data set.
\item AutoMOS~\cite{patton_automos_2016} --- We re-implement the approach, but substitute the synthesized speech embeddings and its auxiliary loss by $L^{\text{MR}}$.
\item Quality-Net~\cite{fu_qualitynet_2018} --- We re-implement the proposed approach.
\item WEnets~\cite{catellier_wenets_2019} --- We adapted the proposed approach to regress MOS.
\item CNN-ELM~\cite{gamper_intrusive_2019} --- We re-implement the proposed approach.
\item NISQA~\cite{mittag_nonintrusive_2019} --- We adapted the proposed approach to work with MOS, and substituted the auxiliary POLQA loss by $L^{\text{MR}}$.
\end{enumerate}

%%%%%%%%%%%%%%%%%%%%%%%%%%%%%%%%%%%%%%%%%%%%%%%%%%%%%%%%%%%%%%%%%%%%%%%%%%%%%%%%%%%%%

\section{Additional Results}
\label{appendix:results}

In Tables~\ref{tab:full_error} and~\ref{tab:full_ablation} we report all error values for the three considered data sets, together with the $L^{\text{TOTAL}}$ average across data sets. Table~\ref{tab:full_error} compares SESQA with existing approaches and Table~\ref{tab:full_ablation} shows the effect of training without one of the considered losses, in addition to using only $L^{\text{MOS}}$. Recall that, as in the main paper, $E^{\text{TOTAL}}=0.5L^{\text{MOS}}+R^{\text{RANK}}+L^{\text{CONS}}$.

\begin{table}[!h]
  \centering
  \begin{tabular}{lcccccccccccccc}
    \hline
    %\multicolumn{2}{c}{\textbf{Ratio}} & 
    %                                     \multicolumn{1}{c}{\textbf{Decibels}} \\
    \textbf{Approach} & & \multicolumn{3}{c}{\textbf{Internal UGC}} & &  \multicolumn{3}{c}{\textbf{Internal VoIP}} & & \multicolumn{3}{c}{\textbf{TCD-VoIP}} & & \textbf{Average} \\
    \cline{3-5} \cline{7-9} \cline{11-13} \\[\dimexpr-\normalbaselineskip+4pt]
       & & \textbf{$L^{\text{MOS}}$} & \textbf{$R^{\text{RANK}}$} & \textbf{$L^{\text{CONS}}$} & & \textbf{$L^{\text{MOS}}$} & \textbf{$R^{\text{RANK}}$} & \textbf{$L^{\text{CONS}}$} & &  \textbf{$L^{\text{MOS}}$} & \textbf{$R^{\text{RANK}}$} & \textbf{$L^{\text{CONS}}$} & & \textbf{$E^{\text{TOTAL}}$} \\
    \hline
    Human & & 0.510 & n/a  & n/a  &
            & 0.755 & n/a  & n/a  &
            & 0.772 & n/a  & n/a  & & n/a \\
    Random score    & & 1.481 & 0.500 & 0.747 &
                    & 1.033 & 0.499 & 0.516 &
                    & 1.144 & 0.501 & 0.580 & & 1.724 \\
    \hline
    ITU-P653        & & 1.304 & 0.501 & 0.050 &
                    & 0.752 & 0.501 & 0.050 &
                    & 0.890 & 0.501 & 0.050 & & 1.042 \\
    FL-JND          & & 0.981 & 0.363 & 0.106 &
                    & 0.768 & 0.411 & 0.078 &
                    & 0.948 & 0.321 & 0.095 & & 0.908 \\
    SRMR            & & 0.995 & 0.283 & 0.110 &
                    & 0.743 & 0.487 & 0.049 &
                    & 0.825 & 0.282 & 0.053 & & 0.849 \\
    FL-PASE         & & 0.798 & 0.291 & 0.126 &
                    & 0.720 & 0.348 & 0.074 &
                    & 0.686 & 0.333 & 0.114 & & 0.796 \\
    \hline
    AutoMOS         & & 0.532 & 0.293 & 0.236 &
                    & 0.536 & 0.292 & 0.250 &
                    & 0.542 & 0.349 & 0.151 & & 0.792 \\
    Quality-Net     & & 0.695 & 0.271 & 0.077 &
                    & 0.657 & 0.319 & 0.075 &
                    & 0.620 & 0.418 & 0.110 & & 0.765 \\
    WEnet           & & 0.702 & 0.211 & 0.142 &
                    & 0.690 & 0.274 & 0.085 &
                    & 0.587 & 0.290 & 0.147 & & 0.713 \\
    NISQA           & & 0.543 & 0.209 & 0.138 &
                    & 0.530 & 0.184 & 0.106 &
                    & 0.594 & 0.335 & 0.125 & & 0.644 \\
    CNN-ELM         & & 0.528 & 0.184 & 0.161 &
                    & 0.511 & 0.176 & 0.130 &
                    & 0.493 & 0.301 & 0.144 & & 0.621 \\
    \hline
    SESQA (ours)    & & 0.485 & 0.096 & 0.089 &
                    & 0.513 & 0.086 & 0.057 &
                    & 0.424 & 0.089 & 0.056 & & 0.394 \\
    \hline
  \end{tabular}
  \caption{Comparison with existing approaches. Error for the considered data sets and metrics.}
  \label{tab:full_error}
\end{table}

\begin{table}[!h]
  \centering
  \begin{tabular}{lcccccccccccccc}
    \hline
    %\multicolumn{2}{c}{\textbf{Ratio}} & 
    %                                     \multicolumn{1}{c}{\textbf{Decibels}} \\
    \textbf{Approach} & & \multicolumn{3}{c}{\textbf{Internal UGC}} & &  \multicolumn{3}{c}{\textbf{Internal VoIP}} & & \multicolumn{3}{c}{\textbf{TCD-VoIP}} & & \textbf{Average} \\
    \cline{3-5} \cline{7-9} \cline{11-13} \\[\dimexpr-\normalbaselineskip+4pt]
       & & \textbf{$L^{\text{MOS}}$} & \textbf{$R^{\text{RANK}}$} & \textbf{$L^{\text{CONS}}$} & & \textbf{$L^{\text{MOS}}$} & \textbf{$R^{\text{RANK}}$} & \textbf{$L^{\text{CONS}}$} & &  \textbf{$L^{\text{MOS}}$} & \textbf{$R^{\text{RANK}}$} & \textbf{$L^{\text{CONS}}$} & & \textbf{$E^{\text{TOTAL}}$} \\
    \hline
    SESQA           & & 0.485 & 0.096 & 0.089 &
                    & 0.513 & 0.086 & 0.057 &
                    & 0.424 & 0.089 & 0.056 & & 0.394 \\
    Without $L^{\text{MOS}}$ & & 1.106 & 0.078 & 0.044 &
                    & 0.700 & 0.074 & 0.044 &
                    & 0.711 & 0.085 & 0.044 & & 0.543 \\
    Without $L^{\text{RANK}}$ & & 0.496 & 0.124 & 0.081 &
                    & 0.544 & 0.277 & 0.050 &
                    & 0.437 & 0.202 & 0.051 & & 0.508 \\
    Without $L^{\text{CONS}}$ & & 0.449 & 0.098 & 0.154 &
                    & 0.464 & 0.086 & 0.117 &
                    & 0.411 & 0.104 & 0.120 & & 0.447 \\
    Without $L^{\text{SD}}$ & & 0.491 & 0.099 & 0.087 &
                    & 0.517 & 0.083 & 0.057 &
                    & 0.437 & 0.090 & 0.057 & & 0.399 \\
    Without $L^{\text{JND}}$ & & 0.484 & 0.096 & 0.089 &
                    & 0.516 & 0.086 & 0.057 &
                    & 0.421 & 0.086 & 0.055 & & 0.394 \\
    Without $L^{\text{DT}}$ & & 0.482 & 0.097 & 0.088 &
                    & 0.523 & 0.082 & 0.056 &
                    & 0.422 & 0.089 & 0.057 & & 0.394 \\
    Without $L^{\text{DS}}$ & & 0.484 & 0.096 & 0.088 &
                    & 0.524 & 0.084 & 0.056 &
                    & 0.429 & 0.089 & 0.057 & & 0.396 \\
    Without $L^{\text{MR}}$ & & 0.500 & 0.104 & 0.086 &
                    & 0.532 & 0.083 & 0.056 &
                    & 0.433 & 0.092 & 0.056 & & 0.403 \\
    Only $L^{\text{MOS}}$ & & 0.478 & 0.208 & 0.163 &
                    & 0.529 & 0.268 & 0.116 &
                    & 0.434 & 0.320 & 0.132 & & 0.643 \\
    \hline
  \end{tabular}
  \caption{Loss ablation study. Error for the considered data sets and metrics.}
  \label{tab:full_ablation}
\end{table}

Fig.~\ref{fig:dists} shows the empirical distribution of distances between latent space vectors $\ve{z}$. We see that smaller distances correspond to similar utterances with the same degradation type and strength (average distance of 7.6 and standard deviation of 3.4), and that larger distances correspond to different utterances with different degradations (average distance of 16.9 and standard deviation of 3.9). The overlap between the two is small, with mean plus one standard deviation not crossing each other. Similar utterances that have different degradations are spread between the previous two distributions (average distance of 13.7 and standard deviation of 5.5). That makes sense in a latent space that is organized by degradation and strengths, with a wide range between small and large strengths. We assume this overall behavior is a consequence of all losses, but in particular of $s$ and $L^{\text{JND}}$ and their linear heads.

\begin{figure}[!h]
  \centering
  \includegraphics[width=0.6\linewidth]{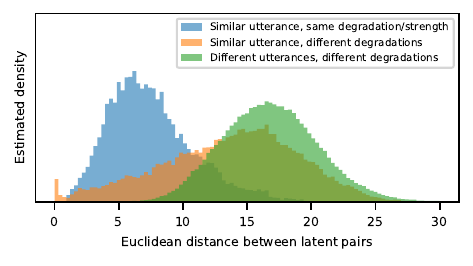}
  \caption{Euclidean distance densities between latent space vectors. Similar/different utterances correspond to the ones defined by $L^{\text{CONS}}$ in the main paper.}
  \label{fig:dists}
\end{figure}

Fig.~\ref{fig:classes} shows three low dimensional t-SNE projections of latent space vectors $\ve{z}$. In the figure, we can see how different degradation types group or cluster together. For instance, with a perplexity of 200, we see that latent vectors of frames that contain additive noise group together in the center. Interestingly, we can also see that similar degradations are placed close to each other. That is the case, for instance, of additive and colored noise, MP3 and OPUS codecs, or Griffin-Lim and STFT phase distortions, respectively. We assume this clustering behavior is a direct consequence of $L^{\text{DT}}$ and its linear head.

\begin{figure}[!h]
  \centering
  \includegraphics[width=0.6\linewidth]{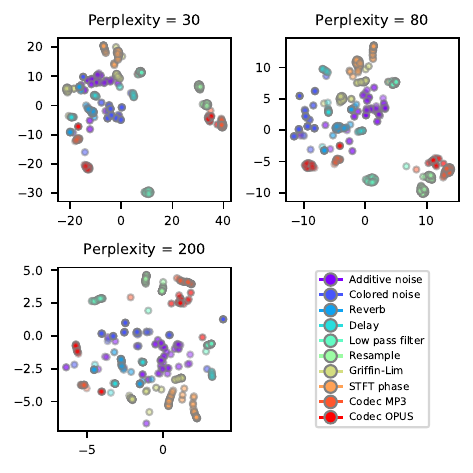}
  \caption{Latent space organization into classes. Projection of $\ve{z}$ by t-SNE with three different perplexities (30, 80, and 200) for a subset of degradation types. Intensity of the color corresponds to degradation strength.}
  \label{fig:classes}
\end{figure}

Fig.~\ref{fig:progdeg} depicts how scores $s$, computed from test signals with no degradation, tend to get lower while increasing degradation strength. In a number of cases, the effect is both clearly visible and consistent (for instance additive noise or the EAC3 codec). In other cases, the effect saturates for high strengths (for instance $\mu$-law quantization or clipping). There are also a few degradations where strength does not correspond to a single variable, and thus the effect is not clearly apparent (for instance, in the reverb degradation, we only control SNR, but there are other variables that also have an important effect in perception). Overall, we observe a consistent behavior across degradations and strengths. We assume $L^{\text{MOS}}$, $L^{\text{RANK}}$, and $L^{\text{DS}}$ are the main driving forces to achieve this behavior.

\begin{figure}[!h]
  \centering
  \includegraphics{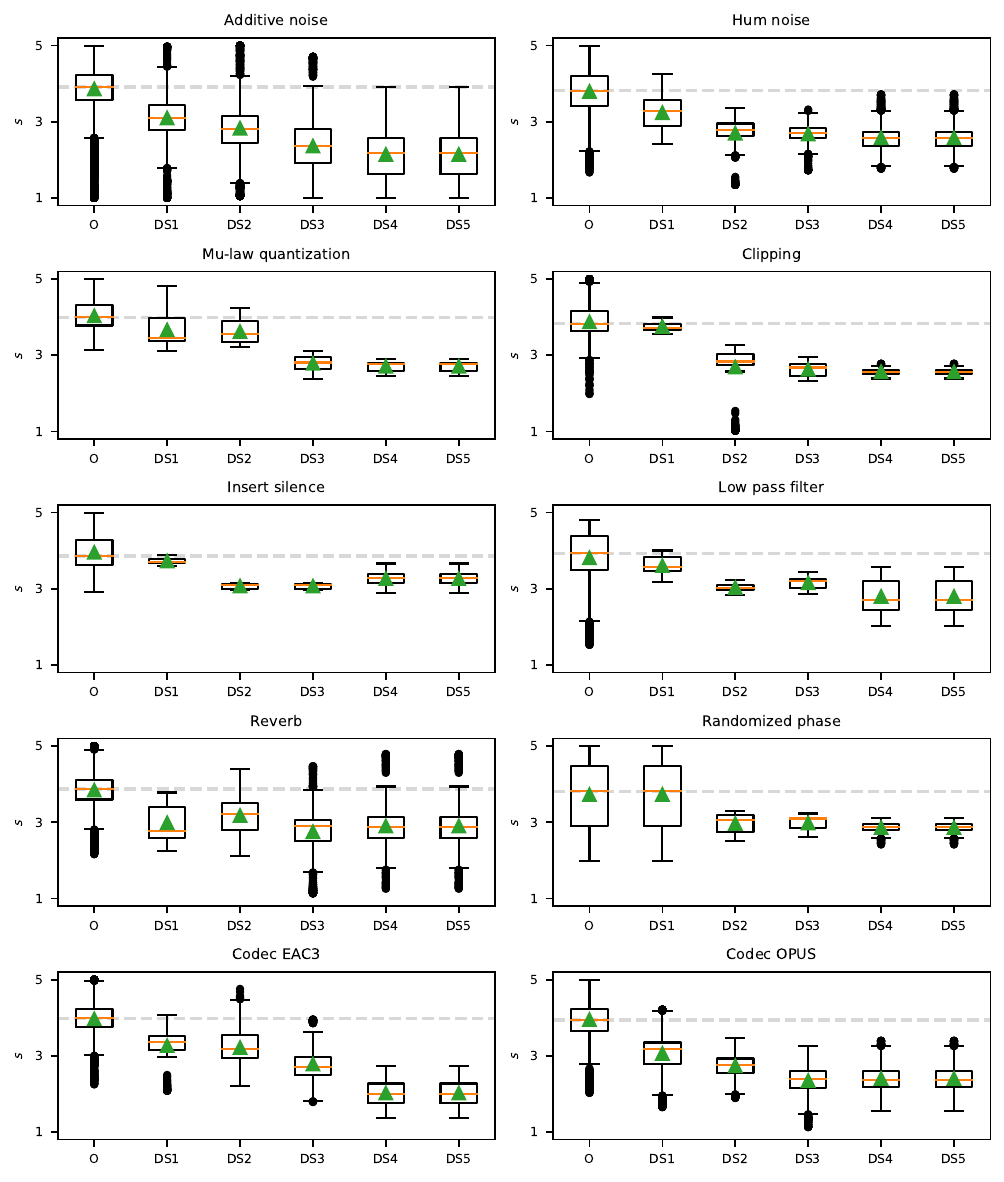}
  \caption{Effect of increasing degradation strength. Original (O) and progressive degradation strength (DS1--5) for a subset of degradation types. Vertical axes correspond to estimated MOS ($s$). The dashed line corresponds to the mean of the original signal score.}
  \label{fig:progdeg}
\end{figure}

%%%%%%%%%%%%%%%%%%%%%%%%%%%%%%%%%%%%%%%%%%%%%%%%%%%%%%%%%%%%%%%%%%%%%%%%%%%%%%%%%%%%%

\section{Code acknowledgment}

In our research we use several codes and scientific programming libraries which we want to acknowledge, the main ones being PyTorch~\cite{paszke_pytorch_2019}, Numpy~\cite{oliphant_guide_2006, van_der_walt_numpy_2011}, Matplotlib~\cite{hunter_matplotlib_2007}, librosa~\cite{mcfee_librosa_2015}, sox~(\url{http://sox.sourceforge.net}), and ffmpeg (\url{http://ffmpeg.org}).

%%%%%%%%%%%%%%%%%%%%%%%%%%%%%%%%%%%%%%%%%%%%%%%%%%%%%%%%%%%%%%%%%%%%%%%%%%%%%%%%%%%%%

\fi

\end{document}